\documentclass[reprint,superscriptaddress,amsmath,amssymb,aps,pra,longbibliography]{revtex4-2}

\usepackage{graphicx}
\usepackage{dcolumn}
\usepackage{bm}
\usepackage[pdftex,breaklinks,colorlinks=true,citecolor=blue,linkcolor=blue,urlcolor=blue]{hyperref}
\usepackage{ulem}
\usepackage{tabularx}
\usepackage{braket}
\usepackage{xcolor}
\usepackage{textcomp}
\usepackage{amsmath}
\usepackage{multirow}
\usepackage{comment}
\usepackage{ragged2e}
\usepackage{natbib}

\begin{document}

\title{High resolution photoassociation spectroscopy of the excited $\cSig$ potential of $^{23}$Na$^{133}$Cs}

\author{Lewis~R. B.~Picard} 
\email{lewispicard@g.harvard.edu}
\affiliation{Department of Chemistry and Chemical Biology, Harvard University, Cambridge, Massachusetts 02138, USA}
\affiliation{Department of Physics, Harvard University, Cambridge, Massachusetts 02138, USA}
\affiliation{Harvard-MIT Center for Ultracold Atoms, Cambridge, Massachusetts 02138, USA}

\author{Jessie~T.~Zhang} 
\altaffiliation[Current affiliation: ]{University of California, Berkeley}
\affiliation{Department of Chemistry and Chemical Biology, Harvard University, Cambridge, Massachusetts 02138, USA}
\affiliation{Department of Physics, Harvard University, Cambridge, Massachusetts 02138, USA}
\affiliation{Harvard-MIT Center for Ultracold Atoms, Cambridge, Massachusetts 02138, USA}

\author{William~B.~Cairncross}
\altaffiliation[Current affiliation: ]{Atom Computing}
\affiliation{Department of Chemistry and Chemical Biology, Harvard University, Cambridge, Massachusetts 02138, USA}
\affiliation{Harvard-MIT Center for Ultracold Atoms, Cambridge, Massachusetts 02138, USA}

\author{Kenneth~Wang} 
\affiliation{Department of Chemistry and Chemical Biology, Harvard University, Cambridge, Massachusetts 02138, USA}
\affiliation{Department of Physics, Harvard University, Cambridge, Massachusetts 02138, USA}
\affiliation{Harvard-MIT Center for Ultracold Atoms, Cambridge, Massachusetts 02138, USA}

\author{Gabriel E. Patenotte} 
\affiliation{Department of Chemistry and Chemical Biology, Harvard University, Cambridge, Massachusetts 02138, USA}
\affiliation{Department of Physics, Harvard University, Cambridge, Massachusetts 02138, USA}
\affiliation{Harvard-MIT Center for Ultracold Atoms, Cambridge, Massachusetts 02138, USA}

\author{Annie J. Park} 
\affiliation{Department of Chemistry and Chemical Biology, Harvard University, Cambridge, Massachusetts 02138, USA}
\affiliation{Harvard-MIT Center for Ultracold Atoms, Cambridge, Massachusetts 02138, USA}

\author{Yichao~Yu} 
\altaffiliation[Current affiliation: ]{Duke University}
\affiliation{Department of Chemistry and Chemical Biology, Harvard University, Cambridge, Massachusetts 02138, USA}
\affiliation{Department of Physics, Harvard University, Cambridge, Massachusetts 02138, USA}
\affiliation{Harvard-MIT Center for Ultracold Atoms, Cambridge, Massachusetts 02138, USA}

\author{Lee~R.~Liu} 
\altaffiliation[Current affiliation: ]{JILA / University of Colorado Boulder}
\affiliation{Department of Chemistry and Chemical Biology, Harvard University, Cambridge, Massachusetts 02138, USA}
\affiliation{Department of Physics, Harvard University, Cambridge, Massachusetts 02138, USA}
\affiliation{Harvard-MIT Center for Ultracold Atoms, Cambridge, Massachusetts 02138, USA}

\author{Jonathan~D.~Hood} 
\affiliation{Department of Chemistry, Purdue University, West Lafayette, Indiana 47907, USA}

\author{Rosario González-Férez}
\affiliation{Instituto Carlos I de Física Teórica y Computacional, and Departamento de Física Atómica, Molecular y Nuclear, Universidad de Granada, 18071 Granada, Spain}

\author{Kang-Kuen~Ni} 
\affiliation{Department of Chemistry and Chemical Biology, Harvard University, Cambridge, Massachusetts 02138, USA}
\affiliation{Department of Physics, Harvard University, Cambridge, Massachusetts 02138, USA}
\affiliation{Harvard-MIT Center for Ultracold Atoms, Cambridge, Massachusetts 02138, USA}

\newcommand{\cSig}{c^3\Sigma_{1}^+}
\newcommand{\cSigDiab}{c^3\Sigma^+}
\newcommand{\aSig}{a^3\Sigma_1^+}
\newcommand{\XSig}{X^1\Sigma^+}
\newcommand{\Na}{{\rm Na}}
\newcommand{\Cs}{{\rm Cs}}
\newcommand{\MHz}{{\rm MHz}}
\newcommand{\GHz}{{\rm GHz}}
\newcommand{\kHz}{{\rm kHz}}
\newcommand{\atomState}{\ket{3,3}_\Cs \ket{1,1}_\Na}
\newcommand{\Mtot}{m_{\rm tot}}
\newcommand{\mW}{{\rm mW}}
\newcommand{\um}{\text{\textmu}{\rm m}}
\newcommand{\us}{\text{{\textmu}s}}
\newcommand{\BPi}{B^1\Pi}
\newcommand{\bPi}{b^3\Pi}

\begin{abstract}
We report on photoassociation spectroscopy probing the $\cSig$ potential of the bi-alkali NaCs molecule, identifying eleven vibrational lines between $v' = 0$ and $v' = 25$ of the excited $\cSig$ potential, and resolving their rotational and hyperfine structure. The observed lines are assigned by fitting to an effective Hamiltonian model of the excited state structure with rotational and hyperfine constants as free parameters. We discuss unexpected broadening of select vibrational lines, and its possible link to strong spin-orbit coupling of the $\cSig$ potential with the nearby $\bPi_1$ and $\BPi_1$ manifolds. Finally we report use of the $v' = 22$ line as an intermediate state for two-photon transfer of weakly bound Feshbach molecules to the rovibrational ground state of the $\XSig$ manifold.
\end{abstract}

\maketitle

\section{Introduction}

Ultracold molecules have in the past two decades been demonstrated to offer a flexible platform for experimental studies of collisions and chemical reaction dynamics \cite{heazlewood_towards_2021, ospelkaus_quantum-state_2010,hu_direct_2019, jurgilas_collisions_2021,yang_evidence_2022,park_feshbach_2023}, many body physics \cite{yan_observation_2013,hazzard_many-body_2014,christakis_probing_2023}, and precision measurement for fundamental physics \cite{hutzler_polyatomic_2020,andreev_improved_2018, kobayashi_measurement_2019, kondov_molecular_2019, roussy_new_2022}. Individually trapped molecules prepared in single quantum states show further promise for quantum information processing \cite{ni_dipolar_2018, sawant_ultracold_2020, yu_scalable_2019, Lin2019, holland_-demand_2022,bao_dipolar_2022}. The realization of these experiments has been enabled by significant advances in methods for preparing and controlling ultracold molecules, using both direct laser cooling and trapping \cite{tarbutt_laser_2018,shuman_laser_2010,anderegg_optical_2019,wu_high_2021,vilas_magneto-optical_2022} and association from ultracold constituent atoms in bulk gases and individual optical traps \cite{Ni2008,yu_coherent_2021,cairncross_rovibgs_2021,he_coherently_2020,reinaudi_optical_2012,takekoshi_ultracold_2014,Molony2014,Guo2016,park_ultracold_2015,stevenson_ultracold_2022}. This rapid proliferation of experimental approaches was facilitated by spectroscopic investigations of relevant molecules, using the tools of atomic and molecular physics. Such investigations have yielded precise measurements of molecular binding energies, rotational constants, Franck-Condon factors, transition dipole moments and higher order perturbations to molecular potentials \cite{mitra_direct_2020,kowalczyk_high_1989,deiglmayr_influence_2009,takekoshi_hyperfine_2011,temelkov_molecular_2015,liu_observation_2016,nakhate_pure_2019}. 

In prior work from our group, we reported on the coherent production of a rovibrational ground state NaCs molecule in an optical tweezer by first magnetoassociating a co-trapped Na and Cs atom pair to a weakly-bound Feshbach molecule followed by performing two-photon Raman transfer to the rovibrational ground state \cite{zhang_forming_2020,cairncross_rovibgs_2021}. NaCs in its rovibrational ground state has one of the largest electric dipole moments of the bi-alkalis (4.6 Debye) \cite{aymar_calculations_2007}, making it a promising candidate for quantum science applications such as quantum computing and quantum simulation. In a separate work from our group, we demonstrated an all-optical approach to coherently transfer from a single pair of atoms in an optical tweezer to a weakly-bound molecule with a binding energy of 770 MHz \cite{yu_coherent_2021}. The successful demonstration of these two-photon molecule creation techniques relied on a good understanding of the excited states that serve as the intermediate in both transfer processes.

\begin{figure}
\centering
\includegraphics[width = 0.95\columnwidth]{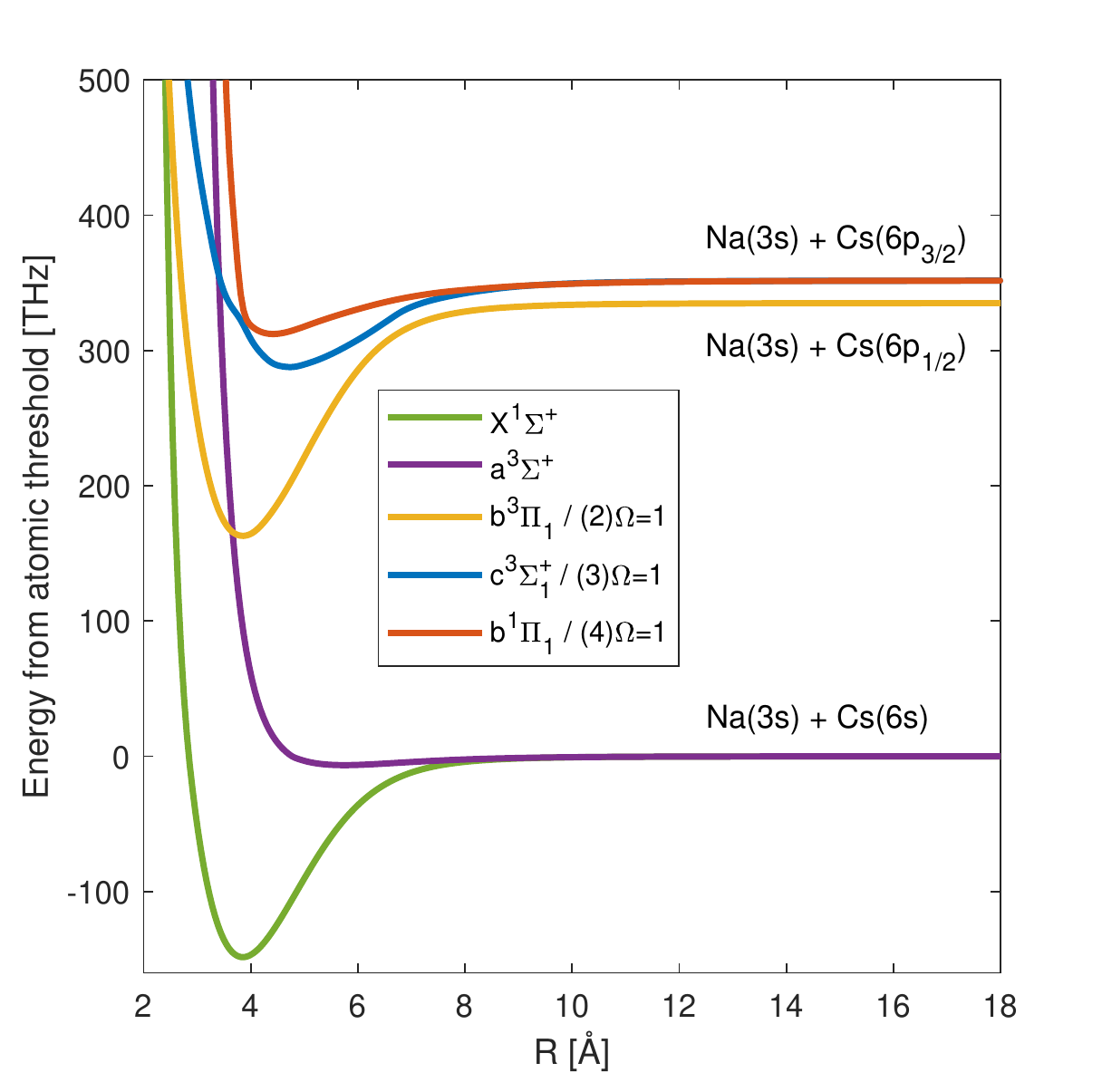}
\caption{Relevant potential energy curves of NaCs. Singlet $\XSig$ (green) and triplet $\aSig$ (purple) curves at the Na(3s) + Cs(6s) threshold are from Ref. \cite{brookes_interaction_2022}. At the two Na(3s) + Cs(6p) thresholds we show empirical potentials for the $\Omega = 1$ components of the $\bPi$ (yellow), $\cSigDiab$ (blue) and $\BPi$ (orange) curves from Refs. \cite{zaharova_solution_2009,grochola_nacspotential_2011,Grochola2010}. These states can also be labelled in the Hund's case (c) basis as $(2)\Omega=1$, $(3)\Omega=1$ and $(4)\Omega=1$, respectively.}
\label{fig:potentials}
\end{figure}

In this work, we present a spectroscopic investigation of the $\cSigDiab_{\Omega=1}$ potential of the NaCs molecule, using direct laser excitation of single atom pairs in optical tweezers. The relevant potential curves of NaCs are shown in Fig. \ref{fig:potentials} \cite{brookes_interaction_2022,zaharova_solution_2009,Grochola2010,grochola_nacspotential_2011}. We observe multiple vibrational lines of the $\cSig$ potential between $v' = 0$ and $v' = 25$. The starting point for photoassociation is a pair of ultracold atoms jointly occupying the ground state of an optical tweezer, from which it is possible to probe the lowest few rotational levels of each vibrational line with rotational and hyperfine resolution. All of the resolved lines are assigned with the aid of an effective Hamiltonian model of the excited state. In previous work, we used the same model Hamiltonian to assign lines in the spectrum of the $v' = 26$ manifold, probed using photon scattering induced loss of a single weakly bound Feshbach molecule \cite{cairncross_rovibgs_2021}.

The paper is organized as follows: in Section \ref{sec:model} we discuss the model of the atomic scattering states, and excited and ground molecular electronic states that are used to fit spectra throughout this work. In Section \ref{sec:csig-spec} we report photoassociation spectroscopy and line assignments of multiple vibrational lines of the excited $\cSig$ potential of NaCs. In Section \ref{sec:linewidths} we discuss our observation of anomalously broad linewidths of states in the $\cSig$--$\bPi_1$--$\BPi_1$ complex and its possible link to the onset of strong coupling between the Hund's case (a) basis states in the complex. Finally in Section \ref{sec:STIRAP} we demonstrate two-photon transfer to the rovibrational ground state using the $\ket{J = 1,m_J=1}$ component of the newly resolved $v' =22$ vibrational state as an intermediate.

\section{Effective Hamiltonian Model}\label{sec:model}

\subsection{$\cSig$ excited state}\label{cSig_model}

The $\cSig$ state of NaCs can be described in the Hund's case (a) basis, as we do here, or in the case (c) basis, in which it would be labelled $(3)\Omega=1$ \cite{Korek2007}. We choose to represent the state in the case (a) basis with quantum numbers
$\ket{\Lambda S \Sigma; J \Omega m_J} \ket{I_\Na m_{I_\Na}} \ket{I_\Cs m_{I_\Cs}}$. Uncoupled nuclear spins are appropriate given the high magnetic fields of 10-860 G used for the spectroscopy reported herein. The high degree of spin-orbit coupling in the \textit{c-b-B} complex splits the $\cSig$ lines from equivalent $c^3\Sigma_{0}^-$ lines, which we have not observed experimentally, by as much as 1 THz \cite{zabawa_production_2012}. We therefore do not consider any terms which mix $|\Omega|=0,1$ states. We model the hyperfine structure of the lowest few rotational levels of $\cSig$ using an effective Hamiltonian including rotational structure, electron spin-nuclear spin hyperfine interactions, Zeeman splitting, and an effective $\Omega$-doubling interaction,
\begin{multline}
    H_{\cSig} = B \, {\bm J}^2 + \alpha_\Na \, {\bm I}_\Na \cdot {\bm S} + \alpha_\Cs \, {\bm I}_\Cs \cdot {\bm S} \\+ g_S \mu_B {\mathcal B} S_z + H_{\Omega}.
\end{multline}
where $B$ is the rotational constant of the molecule, and $\mathcal{B}$ is the applied magnetic field. $\alpha_\text{Na}$ and $ \alpha_\text{Cs}$ are effective hyperfine parameters which describe the strength of each nucleus' magnetic dipole contact interaction with the electron spin. We do not consider quadrupolar hyperfine interactions which, based on the hyperfine splittings of Na and Cs atoms, should be two-to-three orders of magnitude smaller than the dipolar term. The $\Omega$-doubling Hamiltonian matrix elements, $
    \bra{\Omega'} H_{\Omega} \ket{\Omega} = \frac{\omega_{ef}}{2} \, \delta_{\Omega',-\Omega},
$
lift the $\Omega$ degeneracy, causing the eigenstates of $H_{\cSig}$ to be states of good parity $\ket{P=\pm} \sim \ket{\Omega} \pm \ket{-\Omega}$. Because we access only one of these parity states in the experiment the value of $\omega_{ef}$ cannot be determined, so we fix it at an \textit{ad hoc} value of 5 MHz and neglect the dependence of the doubling on $J$. To estimate its Franck-Condon couplings to the Feshbach and $\XSig$ states, we model the vibrational structure of the $\cSig$ state using the experimental potential of Ref.~\cite{grochola_nacspotential_2011}. Given that we only consider states with $|\Omega|=1$ we absorb terms that depend only on $\Omega$ into the overall vibronic energy, such that the line assignment is the same in both the Hund's case (a) and (c) bases.

\subsection{Scattering state}
Following Refs.~\cite{zhang_forming_2020,brookes_interaction_2022}, we use a coupled-channel model for the atom scattering state that includes full details of the $\aSig$ (triplet) and $\XSig$ (singlet) molecular potentials which asymptote to the free atom ground state. This model is designed to describe the Feshbach resonance at 864.11 G used for adiabatic preparation of weakly bound molecules \cite{zhang_forming_2020}, and can predict the bound vibrational and unbound scattering wavefunctions of the atom pair for any combination of initial atomic hyperfine states. The model Hamiltonian consists of Na and Cs hyperfine interactions, $H_{\rm Na,Cs}$, singlet and triplet channel molecular potentials denoted by $V_{\XSig}(R)$ and $V_{\aSig}(R)$ and a magnetic dipole-dipole interaction $V_{d}(R)$:
\begin{multline}
    H_{\rm F}(R) = H_\Na + H_\Cs + {\mathcal P}_{S=0} \, V_{\XSig}(R) \\ + {\mathcal P}_{S=1} \, V_{\aSig}(R) + V_{d}(R),
\end{multline}
where ${\mathcal P_{S=0}}$ and ${\mathcal P_{S=1}}$ are projectors onto singlet and triplet subspaces. We use functional forms of the singlet and triplet potentials and $V_d$ term given in Ref. \cite{brookes_interaction_2022}, with the former being a refinement of potentials originally defined in Ref. \cite{docenko_coupling_2006}. We solve the coupled-channel problem in the spin-coupled basis $\ket{N,m_N; S,m_S; I_\Na,m_{I_\Na}; I_\Cs,m_{I_\Cs}}$, including only channels with no rotational angular momentum, $\ket{N=0,m_N=0}$. Depending on the choice of initial atomic state preparation, we restrict the incoming channels to only those with either $m_{\rm tot}= m_{I_\Na} +  m_{I_\Cs} = 4$ or 6.

In section \ref{sec:STIRAP} of this paper we discuss two-photon transfer to the rovibrational ground state of NaCs in the $\XSig$ potential, building on work in Ref. \cite{cairncross_rovibgs_2021}. We model the rotational and hyperfine structure of the rovibrational ground state using the Hamiltonian and hyperfine constants given in Ref.~\cite{aldegunde_bialkaliHF_2017}. For the vibrational wavefunction we use the same $\XSig$ potential from Ref.~\cite{brookes_interaction_2022} that describes the singlet component of the atomic scattering state.

\begin{figure*}[ht]
\centering
\includegraphics[width = \textwidth]{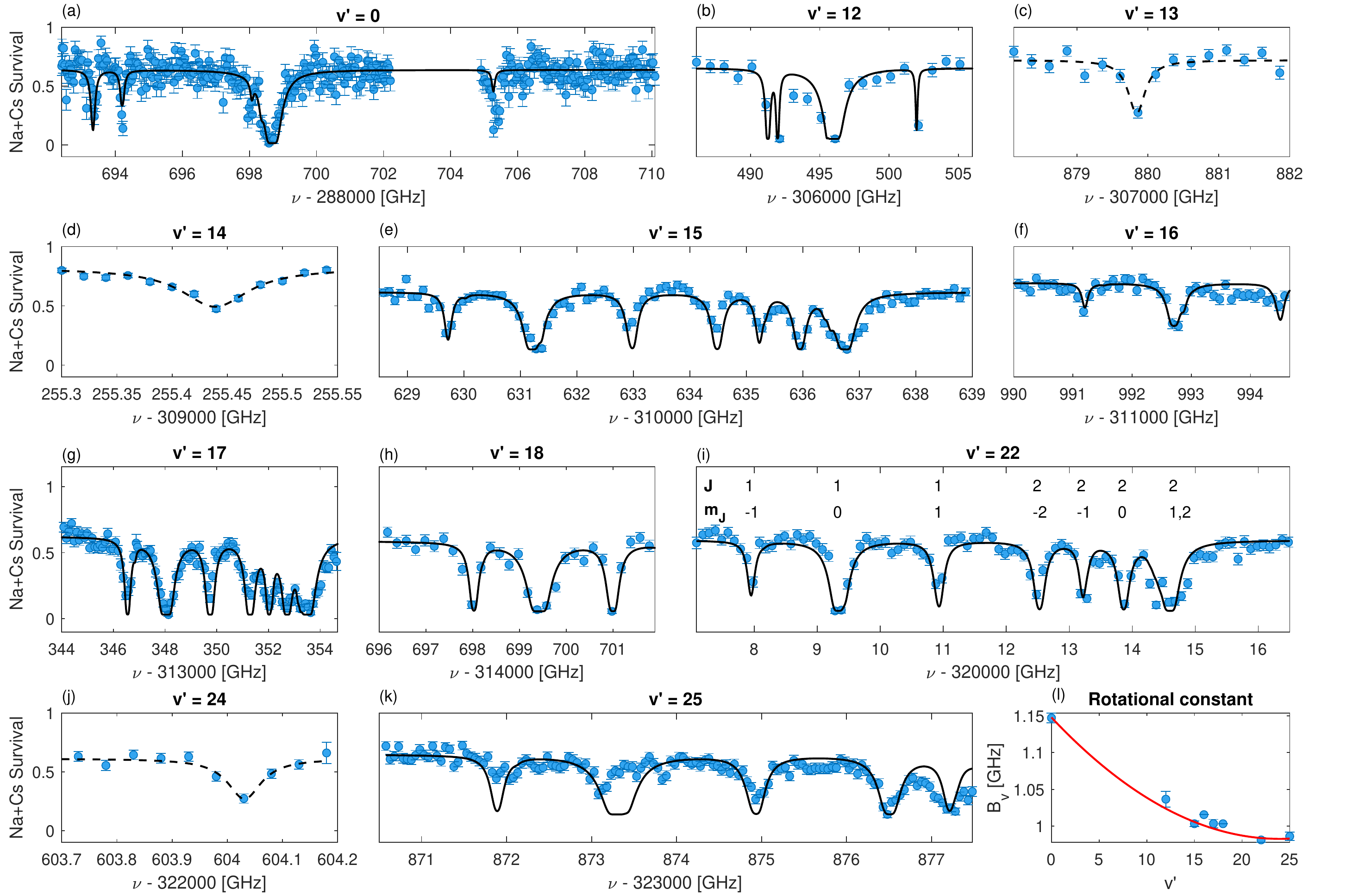}
\caption{\label{fig:paspectrum} (a - k) Photoassociation spectra showing the $v'$ = 0 (a), 12-18 (b - h), 22 (i), 24 (j), and 25 (k) vibrational manifolds of the $\cSig$ potential. The atoms were prepared in $\ket{F=2,m_F=2}_\text{Na}\ket{F = 4, m_F = 4}_\text{Cs}$ ($\ket{F=1,m_F=1}_\text{Na}\ket{F = 3, m_F = 3}_\text{Cs}$) for $v'$ = 0 to 14 ($v'$ = 15 to 25), and a constant magnetic field of $B=8.8$ G ($B=860$ G) was applied. The polarization of the photoassociation light was approximately $\sigma^+$ for $v'$ = 0 to 14, and an equal mixture of $\sigma^+$ and $\sigma^-$ for $v'$ = 15 to 25. Solid lines are fits to the model of the $\cSig$ state described in section \ref{cSig_model} with the rotational constant $B$ and the  hyperfine constant $\alpha_{\textrm{Na}}$ along with a global frequency offset and scattering time as fit parameters. The fitting procedure is discussed in \ref{sec:csig-spec}, and the fit values for each vibrational manifold are given in Table \ref{table:PAFits}. The dashed lines indicate Lorentzian fits, which we use to fit the line position in cases where only a single state is observed without the rotational and hyperfine resolution required to perform a fit to the complete $\cSig$ model. Frequencies of the highest energy line of the $J=1$ manifold of each level are given in Table \ref{table:PALines}. We label the assignment of lines by their approximate $J$ and $m_J$ quantum numbers overlaid on the data for $v' = 22$. (l) Fit of the rotational constant dependence on vibrational level to second order using the model $B_v = B - \alpha(v' + \frac{1}{2}) + \gamma(v' + \frac{1}{2})^2$ \cite{Brown2003}, with parameter fit values $B = 1.154(9)$ GHz, $\alpha = 14(1)$ MHz, $\gamma = 0.29(4)$ MHz.}
\label{fig:SpectroscopyData}
\end{figure*}

\section{Photoassociation spectroscopy}\label{sec:csig-spec}

\begin{table*}
\caption{\label{table:PAFits} Hamiltonian fit parameters of observed lines in the $\cSig$ potential. Levels $v'$ = 13, 14 and 24 are not included, as only one line was observed in each of these vibrational manifolds, which does not provide sufficient data to reliably fit the full $\cSig$ Hamiltonian. In the case of the $v'$ = 12 data we did not measure the spectrum with hyperfine state resolution, so the hyperfine constants were fixed at the fit values for $v'$ = 15 and only the rotational constant $B$ was varied. Conversely, in the case of $v'$ = 16 and 18, only the first rotational level was measured, precluding robust fitting of the rotational constant. For these levels the rotational constant was fixed at the value predicted by the model fit in Fig. \ref{fig:paspectrum}(l) for these vibrational levels.}
\begin{tabularx}{0.95\textwidth}{|X|X|X|X|X|X|X|X|}
\hline
Vibrational state & Initial state $\ket{F,m_F}_\text{Na}$  $\ket{F,m_F}_\text{Cs}$ & Laser pol. & Magnetic field [G] & $B$ [GHz] & $\alpha_{\textrm{Na}}$ [GHz] & Observed rotational lines  \\
\hline\hline
0 & $\ket{4,4}\ket{2,2}$ & $\sigma^+$ & 8.8 & 1.147(7) & 0.69(3) &  $J = 1,2,3$\\
\hline
12 & $\ket{4,4}\ket{2,2}$ & $\sigma^+$ & 8.8 & 1.04(1) & - & $J = 1,2$ \\
\hline
15 & $\ket{3,3}\ket{1,1}$ & $\sigma^+$ + $\sigma^-$ & 860 & 1.003(3) & 0.63(1) &  $J = 1,2$ \\
\hline
16 & $\ket{3,3}\ket{1,1}$ & $\sigma^+$ + $\sigma^-$ & 860 & - & 0.66(3) &  $J = 1$ \\
\hline
17 & $\ket{3,3}\ket{1,1}$ & $\sigma^+$ + $\sigma^-$ & 860 & 1.004(5) & 0.47(5) &  $J = 1,2$ \\
\hline
18 & $\ket{3,3}\ket{1,1}$ & $\sigma^+$ + $\sigma^-$ & 860 & - & 0.40(2) & $J = 1$ \\
\hline
22 & $\ket{3,3}\ket{1,1}$ & $\sigma^+$ + $\sigma^-$ & 860 & 0.981(3) & 0.41(2) &  $J = 1,2$ \\
\hline
25 & $\ket{3,3}\ket{1,1}$ & $\sigma^+$ + $\sigma^-$ & 860 & 0.986(5) & 0.46(2) &  $J = 1$ \\
\hline
\end{tabularx}
\end{table*}

To probe the electronically excited $\cSig$ potential we perform photoassociation (PA) spectroscopy on pairs of Na and Cs atoms co-trapped in a single optical tweezer \cite{hutzler_eliminating_2017}. Both atoms are Raman sideband cooled in separate optical tweezers, which are then adiabatically merged into a single trap \cite{yu_motional-ground-state_2018, liu_molecular_2019}. In this work, the atom hyperfine states are initialized to either $\ket{F=2,m_F=2}_\text{Na}\ket{F = 4, m_F = 4}_\text{Cs}$, with total magnetic quantum number $m_{\rm tot}=6$, or $\ket{F=1,m_F=1}_\text{Na}\ket{F = 3, m_F = 3}_\text{Cs}$, with $m_{\rm tot}=4$. The $m_{\rm tot}=6$ scattering state is probed at a magnetic field of 8.8 G, while the $m_{\rm tot}=4$ state is probed at high magnetic field of 860 G, close to the Feshbach resonance in this channel \cite{zhang_forming_2020,brookes_interaction_2022}.  The magnetic field defines the quantization axis in our system, and is parallel to the k-vector of the tweezer. The atoms are photoassociated using light resonant with the transition from the lowest energy unbound relative motional state of the atoms in the trap to bound molecular states of the $\cSig$ potential. The photoassociation light is provided either by the optical tweezer itself, the frequency of which can be widely tuned while still providing strong confinement of the atoms, or by an additional counter-propagating focused beam overlapping the tweezer. In the former case the tweezer/PA light is generated from a home-built tunable external cavity diode laser, while in the latter the PA light is generated from a tunable Ti:Sapphire laser. When photoassociation occurs the molecular state will generally rapidly decay, either to an excited atomic motional state which is likely to have sufficient energy to escape the trap, or to a lower energy molecular bound state which will be dark to the atom imaging step. We can therefore detect photoassociation through correlated loss of Na and Cs atoms \cite{liu_building_2018}. In the spectra presented here, we plot the joint Na and Cs survival, conditioned on initial loading of both atoms in their respective traps.

We observe photoassociation to the $v' = 0, 12, 13, 14, 15, 16, 17, 18, 22, 24$ and $25$ vibrational manifolds of the $\cSig$ potential. Within some of these vibrational levels we are able to observe rotational structure up to the $J = 3$ rotational state, and perform hyperfine resolved spectroscopy. The spectra of all observed lines in each vibrational level are shown in Fig. \ref{fig:paspectrum} along with corresponding fits to the effective Hamiltonian model. In Table \ref{table:PAFits} we report experimental parameters and fit values for all observed lines, and in Table \ref{table:PALines} we report absolute frequency measurements and linewidths of the $\ket{J=1,m_J=1}$ hyperfine state for each vibrational level. We choose this line as a reference because it is a resolved state to which we can consistently strongly couple with pure $\sigma^+$ or  $\sigma^+ + \sigma^-$ light (defined with respect to the magnetic quantization axis) at either high or low magnetic  field, and because it is the intermediate state which we use for two-photon transfer to the rovibrational ground state, discussed in Section \ref{sec:STIRAP} and in Refs. \cite{cairncross_rovibgs_2021,zhang_optical_2022}.

For the initial atom pair state $\ket{F=2,m_F=2}_\text{Na}\ket{F = 4, m_F = 4}_\text{Cs}$, the angular momentum values are predominantly $N=0,~ S=1$, giving $J=1$. By selection rules, this state couples most strongly to $J=2$ states in the excited state manifold with predominantly $\sigma^+$ polarization, with a small coupling to $J= 3$ due to mixing of rotational states in $\cSig$. Experimentally we observe stronger than expected coupling to $J= 3$, however, which we attribute to imperfect motional cooling of the atom pair, which would lead to having some initial population in a rotationally excited atom pair state which couples more strongly to $J = 3$.

On the other hand, the initial atom pair state $\ket{F=1,m_F=1}_\text{Na}\ket{F = 3, m_F = 3}_\text{Cs}$ is in a mixed singlet-triplet spin state, and can thus couple to both $J=1$ and $J=2$ via  $\sigma^+$ polarized light. This scattering channel exhibits a Feshbach resonance at 860.11(5) G, which we use for adiabatic assembly of weakly bound Feshbach molecules \cite{zhang_forming_2020}. Part of the motivation for performing the spectroscopy in this work is to identify vibrational states in $\cSig$ suitable to use as an intermediate state for two-photon optical transfer from these Feshbach molecules to the rovibrational ground state. In Ref.~\cite{cairncross_rovibgs_2021} we showed spectra resolving the hyperfine structure of the $v' = 26$ level with $\sigma^+ + \sigma^-$ light incident on Feshbach molecules, and used the $\ket{J=1, m_J = 1}$ line of this vibrational state as such an intermediate to prepare rovibrational ground state NaCs molecules via a detuned Raman process. Our choice of two-photon transfer method and transfer efficiency were constrained by the anomalously broad linewidth of $v' = 26$, discussed further in Section \ref{sec:linewidths}. We presently use the $\ket{J=1, m_J = 1}$ line in the $v' = 22$ level identified in this work for ground-state transfer, described in Section \ref{sec:STIRAP}.

For each vibrational level, we use the scattering and excited state models described in Section \ref{sec:model} to fit the observed spectra. Our initial state is fixed to be either the $\ket{F=2,m_F=2}_\text{Na}\ket{F = 4, m_F = 4}_\text{Cs}$ state at low field or the $\ket{F = 1, m_F = 1}_\text{Na}\ket{F=3,m_F=3}_\text{Cs}$ state at high field, with the vibrational wavefunctions corresponding to each hyperfine component of these scattering channels determined by the coupled-channel model. Because we perform Raman sideband cooling on both atoms before PA, preparing each atom in its absolute motional ground state with high probability \cite{yu_motional-ground-state_2018,liu_ground_2018,zhang_assembling_2021}, we assume that the initial scattering state is the lowest energy unbound eigenstate found when solving the coupled-channel problem. The forms of the spectra obtained for a given initial state are determined by a combination of the vibrational wavefunctions and the spin and angular momentum expansion factors for each of the available excitation channels. The transition dipole moment (TDM) to every excited state is found by first integrating the scattering wavefunction over internuclear separation \textit{R}, multiplied by a function from \textit{ab initio} theory describing the position dependence of the singlet-triplet character of the excited state \cite{rosario_private,yu_coherent_2021} and then summing over the components of the scattering wavefunction weighted by their respective angular momentum coupling coefficients. 

In fitting the spectra we vary the rotational constant $B$ and the hyperfine constant $\alpha_{\textrm{Na}}$  as free parameters. We find that if both  $\alpha_{\textrm{Cs}}$ and $\alpha_{\textrm{Na}}$ are used as free parameters they exhibit a high degree of covariance, such that neither can be accurately uniquely determined by fitting the spectra. Given that the $\cSig$ potential asymptotes to the Na(3s) + Cs(6p$_{3/2}$) pair, the Na ground state hyperfine splitting will be the dominant contribution to the molecule's hyperfine structure. As such, we approximate the $\alpha_{\textrm{Cs}}$ parameter by its 6p$_{3/2}$ asymptotic value of 50.275 MHz \cite{steck_cs}. We note that this may be an overestimate of the Cs contribution, and thus the Na hyperfine constants in Table \ref{table:PAFits} may represent a slight underestimate of the Na contribution to the hyperfine structure. We additionally use global frequency offset and scattering time variables to simultaneously fit all the line positions and the magnitude of the PA depletion signal, respectively. At each fit iteration, we diagonalize the excited state Hamiltonian to determine line positions and relative strengths, from which we derive the expected atom pair survival as a function of PA frequency. We perform a nonlinear least squares fit to minimize the difference between the simulated and observed spectra simultaneously for all four free parameters. Fit parameters for each vibrational level are given in Table \ref{table:PAFits}. With the aid of the model we can identify each of the lines, which are labeled, overlaid on the $v' = 22$ spectrum in Fig. \ref{fig:paspectrum}(i), by $J$ and $m_J$, which are approximately good quantum numbers for the state. The ordering of lines as a function of energy is the same for all of the vibrational levels. At the lower magnetic field used to probe $v'$ = 0 to 14 the hyperfine substructure of each rotational level is not fully resolved. In Fig. \ref{fig:paspectrum}(l) we fit the vibrational dependence of the rotational constant of the molecule to second order in $v'$ using the functional form $B_v = B - \alpha(v' + \frac{1}{2}) + \gamma(v' + \frac{1}{2})^2$ \cite{Brown2003} with parameter fit values $B = 1.154(9)$ GHz, $\alpha = 14(1)$ MHz, $\gamma = 0.29(4) $MHz. We note also that there is a clear decrease in the effective Na hyperfine coupling constant $\alpha_{\textrm{Na}}$ as a function of $v'$, indicating a reduction of electron density at the Na nucleus at higher vibrational levels \cite{kowalczyk_high_1989}.

\begin{table}
    \centering
    \begin{tabularx}{\columnwidth}{l l l}
        \hline \hline
        $v'$ & $\ket{J=1,m_J = 1}$ freq [GHz] & Linewidth [MHz] \\ \hline
        0 & 288698.91(5) & 39(13)\\
        12 & 306496.6(1) & 27(1) \cite{yu_coherent_2020}\\
        $13^*$ & 307884.55(8) & -\\
        $14^*$ & 309260.150(4) & 70(10) \\
        15 & 310624.77(1) & 12(3) \\
        16 & 311986.30(2) & 10(3) \\
        17 & 313341.27(2) & 12(8) \\
        18 & 314692.77(1) & - \\
        22 & 320002.49(1) & 17(3) \\
        $24^*$ & 322595.86(1) & - \\
        25 & 323866.47(8) & 42(9) \\
        26 & 325121.31(7) & 120(30) 
    \cite{cairncross_rovibgs_2021} \\
    \end{tabularx}
    \caption{\label{table:PALines} Frequencies and narrowest observed linewidths of $\ket{J=1,m_J = 1}$ states in each vibrational level. For levels $v'$ = 13, 18 and 24 the lines were only observed in the presence of significant power broadening, so we do not report a narrow linewidth for these states. * denotes levels for which coupling to the state was weak enough that only one line could be observed experimentally, making rotational and hyperfine assignment impossible. For these levels, the reported frequency is the center of a Lorentzian fit to the observed line. For all other lines, the frequency of the $\ket{J=1,m_J = 1}$ line is determined by fitting the whole spectrum to the $\cSig$ model as described in the text. The reported transition frequencies are given with respect to the hyperfine center of mass of the atom pair at the magnetic field used for spectroscopy \cite{note_hfCOM}. The frequency and linewidth of the $v' = 26$ state, previously reported in \cite{cairncross_rovibgs_2021}, were determined using resonant depletion of Feshbach molecules, are included here for completeness.}
    \label{tab:fb_state_admixture}
\end{table}

\section{Anomalous broadening in the cbB-complex}\label{sec:linewidths}
A distinctive feature of the vibrational spectra of the $\cSig$ potential is the presence of several vibrational lines with anomalously broad linewidths. Similar anomalous broadening of lines in the coupled $\cSig-\bPi_{1}-\BPi_{1}$ complex has previously been observed in other bialkalis, including Li$_2$\cite{schmidt-mink_predissociation_1985}, NaK \cite{temelkov_molecular_2015} and NaRb \cite{guo_high-resolution_2017}, but the origin of the broadening remains poorly understood. In Table \ref{table:PALines}, we report measurements of the natural linewidths of the $\ket{J=1,m_J = 1}$ line for most of the vibrational states identified in this work. In order to measure the natural linewidth without power broadening, we systematically lower the intensity of the photoassociation laser until only approximately 50\% of the atom population is photoassociated. We then scan the laser frequency across the line and perform a fit to a Lorentzian to determine the linewidth. In Ref. \cite{cairncross_rovibgs_2021} we investigated only the $v'$ = 26 vibrational level with the goal of using it as an intermediate for ground state molecule production, and found it to have a natural linewidth of 120(30) MHz, more than an order of magnitude larger than the Cs D2 line to which the molecular potential asymptotes, which has a full width at half maximum of 5.2 MHz \cite{steck_cs}. We note that independent work in Ref.~\cite{stevenson_ultracold_2022} recently measured this same state to have a linewidth of 51(5) MHz, probed in a bulk sample of NaCs in a weak optical dipole trap. In our spectroscopic investigation we identify several  narrower vibrational lines, albeit with the narrowest having a linewidth of 10 MHz, still a factor of 2 larger than the Cs D2 line. The narrowest lines we observe are $v'$ = 15-17, with $v'$ = 22 having a slightly larger linewidth of 17 MHz and $v'$ = 25 and 26 both being significantly broader with linewidths of 42(9) and 120(30) MHz, respectively.

\begin{figure} [h]
\centering
\includegraphics[width = 0.9\columnwidth]{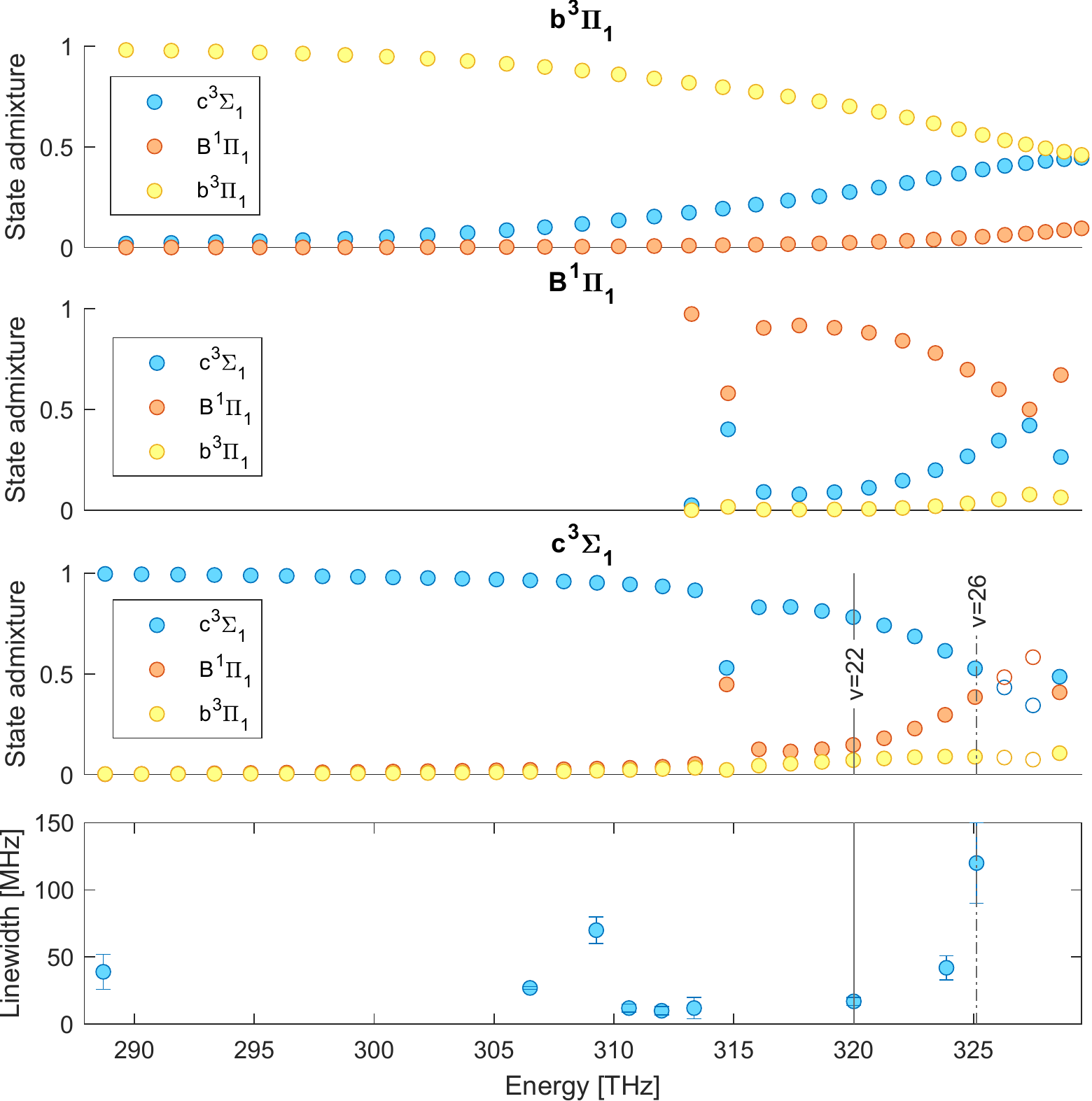}
\caption{Predicted admixtures for each state as a function of energy from the Na(3s) + Cs(6s) threshold, shown from the bottom of the $\cSig$ potential to just below the Na(3s) + Cs(6p$_{3/2}$) threshold. Particular eigenstates of the coupled potential are assigned to the uncoupled potential of which they have the largest admixture, with the exception of highly mixed states denoted by hollow circles, which are plotted for clarity with the $\cSig$ vibrational progression to which they belong, though they have a slightly larger admixtures of $B^1\Pi$. Also shown in the bottom panel are the experimentally observed linewidths, reported in Table \ref{table:PALines}, for all the $\cSig$ states identified in this work.}
\label{fig:admixtures}
\end{figure}

One of the possible mechanisms for the observed broadening is predissociation, coupling of a molecular state to another state at an energy above the latter's dissociation threshold, which can lead to rapid nonradiative dissociation of the molecule \cite{Brown2003_predissoc,zabawa_production_2012}. This has previously been proposed as a potential mechanism for the anomalous broadening observed in  Li$_2$ \cite{schmidt-mink_predissociation_1985} and NaRb \cite{guo_high-resolution_2017}. In NaCs, the $\cSig$ potential is mixed very strongly with the nearby $\BPi_1$ and $\bPi_1$ potentials via spin-orbit coupling. We note that while earlier widely used experimental potentials for NaCs \cite{docenko_coupling_2006} predict that the $\aSig$ curve does not cross any electronically excited potentials below the Cs(6p) threshold, the recent refinement of the triplet scattering potential \cite{brookes_interaction_2022} shown in Fig. \ref{fig:potentials} exhibits a significantly sharper short-range repulsive wall which crosses the $\bPi_1$ curve well below threshold. This is the same potential which we use in the coupled-channel model and which accurately reproduces the location of the Feshbach resonance in the lowest energy scattering channel. Though the precise short range behavior may be difficult to predict with confidence from scattering data, this new prediction suggests that there may be lower energy crossings and more significant wavefunction overlap between bound and dissociative states than previously thought. 

While predissociation remains difficult to observe directly, we seek to gain some understanding of the potential role of this mechanism in the broadening of some vibrational lines by modelling the nonradiative couplings of the $\cSig$ potential to $\BPi_1$ and $\bPi_1$. We take as our starting point previous \textit{ab initio} calculations of the spin-orbit couplings between the states in the $c-b-B$ complex as a function of internuclear separation $R$ \cite{rosario_private,skomorowski_rovibrational_2012}, which capture the overall coupling but do not accurately reproduce experimental spectra of the molecule. We then add $R$-dependent perturbations of the form $W(R) = c~(R/R_0)^n\exp(-(R/R_0)^m)$ to the off-diagonal coupling matrix elements and the diagonal terms representing the \textit{ab initio} potentials, and optimize the parameters $c$, $R_0$, $n$ and $m$ for each term to minimize the residual between the deperturbed potential and empirical adiabatic $\cSig$, $\bPi_1$ and $\BPi_1$ potentials \cite{zaharova_solution_2009,Grochola2010,grochola_nacspotential_2011}. This process is known as deperturbation \cite{zaharova_solution_2009,kotlar_analysis_1980}. Using the deperturbed potentials, we can then estimate the energies and admixtures of each of the three states for all of the vibrational levels in the  $c-b-B$ complex, as shown in Fig. \ref{fig:admixtures}. Note that, because the mixing depends significantly on the spacings between the closest vibrational levels from each series in the complex, the deperturbation step is very important for capturing trends in the mixing strength as a function of energy. We chose the deperturbation matrix element to have a relatively small number of free parameters and a simple functional form which goes to 0 at $R = 0,\infty$. However, the deperturbation elements themselves are phenomenological. With this caveat in mind, we note that the deperturbed potentials predict a very strong mixing of all three levels, particular $\cSig$ and $\BPi_1$, in the region between $v'$ = 23 and 32 of the $\cSig$ vibrational series. The most mixed states, $v'$ = 27 and 28 of $\cSig$, are actually close to an even mixture of singlet and triplet, but for simplicity's sake we still label them as part of the $\cSig$ vibrational series. This region of predicted strong mixing coincides with our experimentally observed onset of anomalous line broadening, suggesting that the mixing may be responsible for inducing rapid predissociation or other nonradiative decay of the states. While our experiment is not equipped to study the potential predissociation process directly, these results indicate that further study of nonradiative decay processes in NaCs and other bialkalis  may prove fruitful.

\section{Two-photon transfer to $\XSig$}\label{sec:STIRAP}
\begin{figure}[ht]
\centering
\includegraphics[width = 0.85\columnwidth]{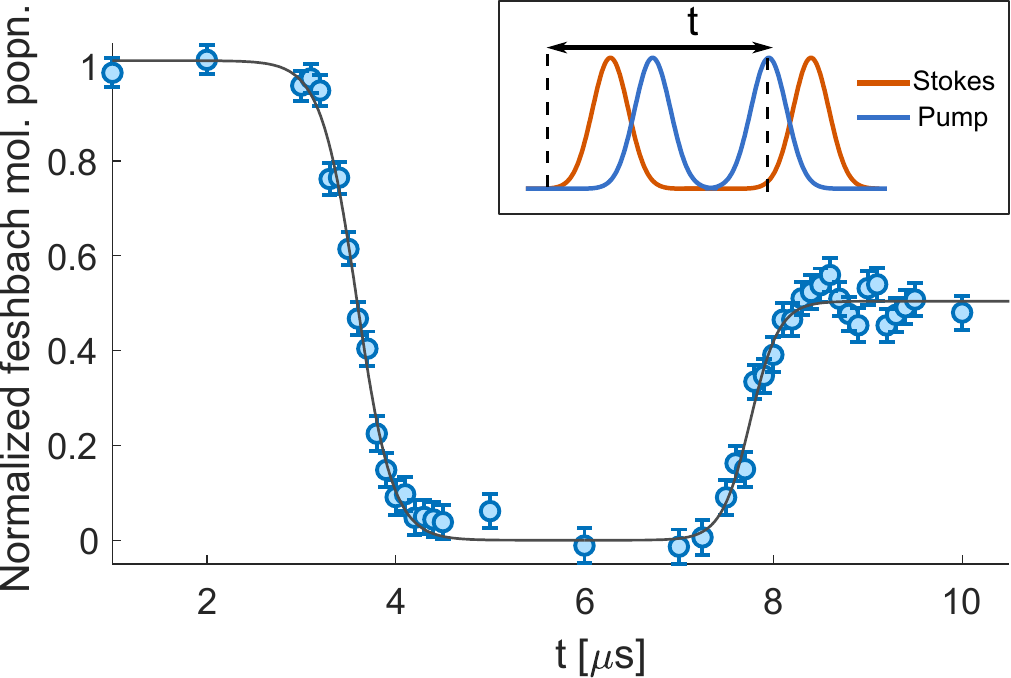}
\caption{STIRAP transfer from Feshbach molecules to the rovibrational ground state of $\XSig$, via the $\cSig, v' = 22$ intermediate state, averaged over eight sites in a 1D optical tweezer array. The pump laser wavelength is 937 nm and the Stokes laser wavelength is 642 nm. The average one-way transfer efficiency is 73(3)\%.}
\label{fig:STIRAP}
\end{figure}

In previous work \cite{cairncross_rovibgs_2021,zhang_optical_2022} we prepared up to five NaCs molecules in an optical tweezer array in the rovibrational ground state of the $\XSig$ potential through Feshbach association followed by detuned two-photon Raman transfer, using the $v' =26$ level of $\cSig$ as an intermediate state. We chose to use detuned Raman transfer because the large linewidth of the $v' = 26$ state (see \ref{table:PALines}) made the more standard resonant STImulated Raman Adiabatic Passage (STIRAP) technique prohibitively lossy. Detuned Raman transfer can minimize scattering from the intermediate, but remains difficult to further scale to a larger array of molecules because the two-photon resonance position is highly sensitive to local variations in the intensity of the Raman beams. Our first demonstration of parallel production of an array of ground state molecules was limited by a combination of scattering from the intermediate state and residual non-uniformity in our top-hat shaped Raman beams \cite{zhang_optical_2022}.

Having now identified several narrower vibrational lines in the $\cSig$ manifold, we select the $v' =22$ level as the new intermediate for ground state transfer. While it does not have the narrowest linewidth of all the lines observed, it is sufficiently close to $v' = 26$ to allow locking of our transfer lasers (pump laser at 937 nm, Stokes laser at 642 nm) to the same high finesse cavity (Stable Laser Systems) that we used previously, and also has similar couplings strengths as $v' = 26$ to both our initial Feshbach molecule state and the rovibrational ground state. Using this new intermediate state, we are able to simultaneously transfer eight molecules to the rovibrational ground state via STIRAP. We use Rabi frequencies of $2\pi \times 40$ MHz for the Stokes laser and  $2\pi \times 14$ MHz for the pump laser, and perform resonant STIRAP ($\Delta \approx 0$) with an average one-way transfer efficiency of 73(3)\%. The NaCs Feshbach molecule population during forward and reverse STIRAP pulses is shown in Fig. \ref{fig:STIRAP}. The number of molecules in the array is currently limited by available trapping laser power at the Na Raman sideband cooling stage, prior to molecular assembly \cite{yu_motional-ground-state_2018}. As resonant STIRAP is highly insensitive to local site-by-site intensity variations, this transfer could be straightforwardly extended to a larger array with more trapping laser power. 

\section{Conclusion}\label{sec:conclusion}

In this work, we performed photoassociation spectroscopy on Na and Cs atom pairs in optical tweezers to probe the $\cSig$ excited state potential and reported rotational and hyperfine-resolved spectra for eleven vibrational states. By using an effective Hamiltonian model of the excited state we assigned the observed lines in the spectra and estimated rotational and hyperfine constants of the state for each vibrational level. We highlighted a possible mechanism for the as-of-yet unexplained broadening of excited state vibrational lines in NaCs and other bi-alkalis. Further study of this broadening may elucidate the complex excited state structure of these molecules and facilitate future experiments which use electronically excited states as intermediates for coherent preparation of ground state molecules. This spectroscopic work allowed us to perform such a two-photon coherent transfer from atom pairs to weakly-bound molecules \cite{yu_coherent_2021} and from weakly-bound molecules to rovibrational ground state molecules in optical tweezers \cite{cairncross_rovibgs_2021}. The coherent creation of molecules in optical tweezers allows for control over all their internal and external degrees of freedom, providing a rich resource for many quantum science applications.

\begin{acknowledgments}
We thank Jeremy Hutson for valuable discussions and feedback on this work and Robert Moszynski for sharing the \textit{ab initio} matrix elements used in section \ref{sec:linewidths}. This work is supported by the NSF (PHY-2110225), the AFOSR (FA9550-19-1-0089), and the Camille and Henry Dreyfus Foundation (TC-18-003). J. T. Z. was supported by a National Defense Science and Engineering Graduate Fellowship. K. W. was supported by an NSF GRFP fellowship. R. G. F. acknowledges financial support by the Spanish projects PID2020-113390GB-I00 (MICIN), PY20-00082 (Junta de Andalucía) and A-FQM-52-UGR20 (ERDF-University of Granada), and the Andalusian Research Group FQM-207.
\end{acknowledgments}

\appendix*
\section{Deperturbation matrix elements}
\begin{figure}[ht]
\centering
\includegraphics[width = 0.85\columnwidth]{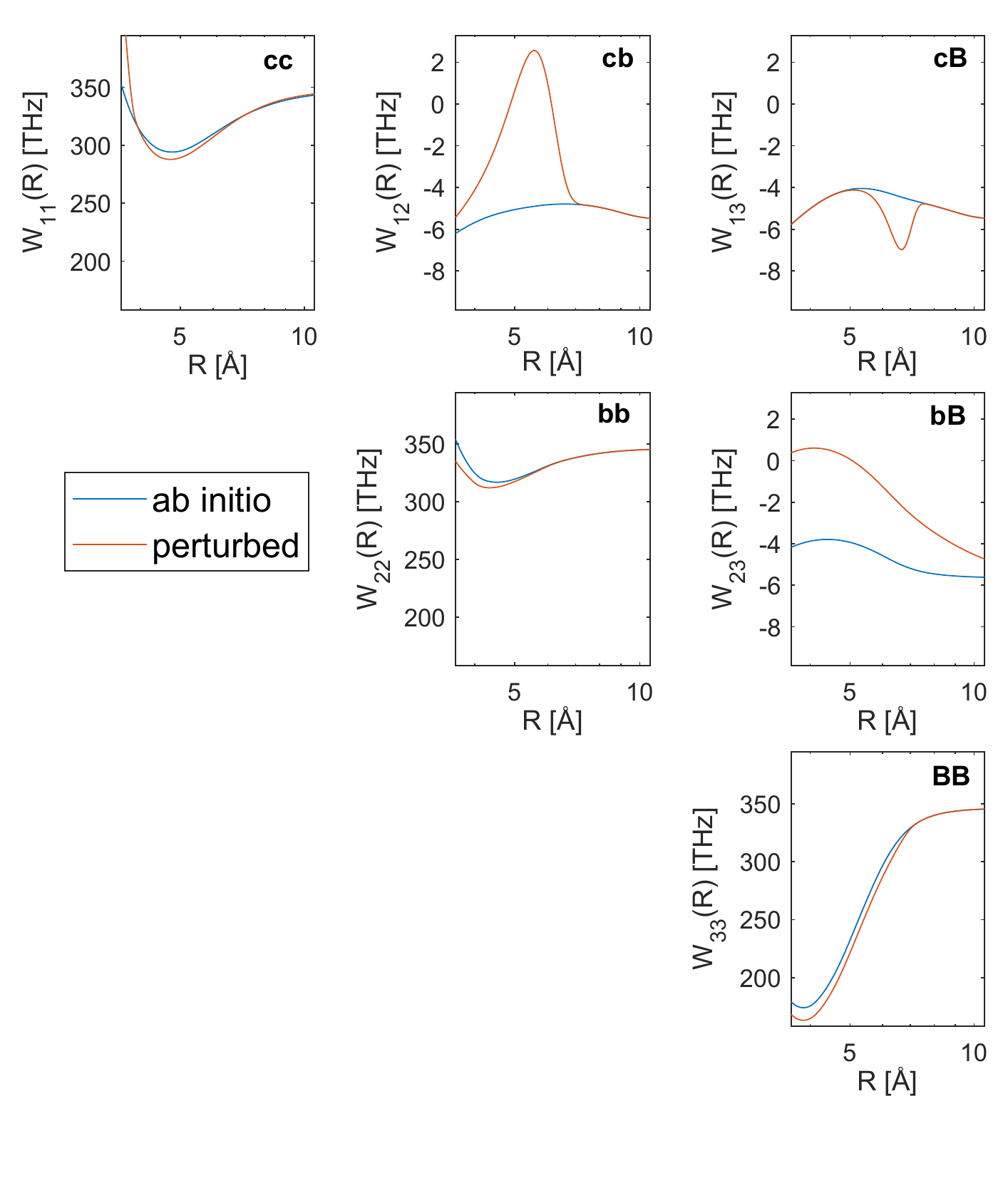}
\caption{\textit{Ab initio} and perturbed matrix elements, in THz, as a function of internuclear separation $R$ for each channel in the coupled $c-b-B$ complex.}
\label{fig:depertMEl}
\end{figure}
In the main text we describe the deperturbation procedure applied to \textit{ab initio} potentials of the $\cSig$, $\bPi_1$ and $\BPi_1$ states to achieve good agreement with empirical curves for these states. We report in Table \ref{table:depertTab} the optimized parameters for the deperturbation functions that we add to each matrix element in the 3x3 coupled-channel Hamiltonian representing the $c-b-B$ complex. In Fig. \ref{fig:depertMEl} we show the \textit{ab initio} and perturbed diagonal potentials and coupling matrix elements used to estimate the mixing between the channels in Section \ref{sec:linewidths}.
To find the optimal deperturbation parameters, we performed an unconstrained search using MATLAB to minimize the residual between the deperturbed potential and the empirical potentials from Refs. \cite{zaharova_solution_2009,Grochola2010,grochola_nacspotential_2011} as a function of all the perturbing function parameters, in the range from $R = $ 3.6 to 10.6 \AA. The exponent parameters $n$ and $m$ are bounded to be greater than 0 to ensure the perturbations go to 0 at $R = 0,\infty$, and all other terms are unbounded. We began by including six perturbing functions, one for each of the diagonal, cc, bb, and BB, and coupling, cb, cB and bB, matrix elements. Following initial optimization, we added additional perturbing functions until reasonable agreement with the empirical potentials was achieved. After optimization we found that the vibrational energies predicted using the perturbed potentials agreed with all the observed lines reported in Table \ref{table:PALines} with a mean absolute deviation of $< 40$ GHz. We note that given the high dimensionality of the minimization problem, the parameters reported here likely do not represent a unique set of optimal deperturbing functions. However, we observe that the energy dependence of the mixing between the diabatic states shown in Fig. \ref{fig:admixtures} is relatively insensitive to the precise form and relative weights of the deperturbing functions, but very sensitive to the spacings between particular vibrational levels. For this reason, we believe that any deperturbed model of the complex that is able to achieve close agreement with experimental data is sufficient to provide qualitative insight into the coupling and its potential link to the widths of the vibrational lines observed herein. We found that using a 2x2 coupled-channel model including only the $\cSig$ and $\BPi_1$ states we could not achieve better than 1 THz agreement with experiment regardless of the number of deperturbing parameters used. While an expanded coupled-channel model involving all of the $\Omega = 0,~1$ and 2 states dissociating to the Cs(6p) thresholds is beyond the scope of this work, it would potentially allow for more precise predictions of the properties of individual vibrational lines. 

\begin{table}[h]
\begin{tabularx}{0.9\columnwidth}{l l l l l}
\hline 
Chan. & $c$ [THz] & $R_0$ [Å] & $n$ & $m$ \\ 
\hline 
\hline 
cc & 4032 & 2.912 & 2.057[-4] & 7.29 \\ 
cc & -26.56 & 3.809 & 2.272 & 2.3 \\ 
cc & 53.09 & 3.748 & 0.4164 & 42.09 \\ 
cc & 65.74 & 3.448 & 0.01794 & 5.972 \\ 
cc & 14.63 & 0.5263 & 1.484[-12] & 0.3209 \\ 
bb & -28.97 & 3.442 & 9.359[-3] & 11.67 \\ 
bb & -146.8 & 1.978 & 0.6053 & 1.69 \\ 
BB & -10.91 & 6.687 & 5.156[-3] & 19.89 \\ 
bB & 9.128 & 1.834 & 1.836 & 0.9779 \\ 
cB & -6.4 & 6.49 & 19.66 & 14.18 \\ 
cb & 19.39 & 5.753 & 6.855 & 9.59 \\ 
\end{tabularx}
\caption{Optimized parameters of deperturbation functions giving perturbed matrix elements shown in Fig. \ref{fig:depertMEl}. Where used, the numbers in square brackets represent the power of 10 by which the preceding parameter is divided. Parameters names are defined in Section \ref{sec:linewidths}.}
\label{table:depertTab}
\end{table}

\bibliographystyle{apsrev4-2}
\bibliography{NaCsSpecBib}

\begin{thebibliography}{69}
\bibitem [{not()}]{note_hfCOM}%
  \BibitemOpen
  \href@noop {} {}\bibinfo {note} {The hyperfine COM was calculated using the
  Breit Rabi formula. It is 4.711 GHz below the
  $\ket{F=2,m_F=2}_\text{Na}\ket{F = 4, m_F = 4}_\text{Cs}$ channel at 8.8 G
  and 8.175 GHz above the $\ket{F=1,m_F=1}_\text{Na}\ket{F = 3, m_F =
  3}_\text{Cs}$ channel at 860 G.}\BibitemShut {Stop}%
\end{thebibliography}

\end{document}